\documentclass[a4paper,11pt]{article}
\pdfoutput=1

\usepackage{jcappub,xcolor} 

\usepackage[T1]{fontenc}
\usepackage[utf8]{inputenc}

\usepackage{bm}
\usepackage{footnote}
\usepackage{latexsym,epsfig,amssymb, amsmath, nicefrac, slashed, color, graphicx,amsfonts}
\usepackage{ulem}
\usepackage{hyperref}
\usepackage{mathtools}
\usepackage{verbatim}

\pdfoutput=1 
\newcommand{\ud}{\,\mathrm{d}}

\usepackage{bbold}

\usepackage{float}

\renewcommand{\bf}[1]{\textbf{#1}}

\title{Many-field Inflation: Universality or Prior Dependence?}

\author[a]{Perseas Christodoulidis,}
\author[a]{Diederik Roest,}
\author[b]{Robert Rosati}
\affiliation[a]{Van Swinderen Institute for Particle Physics and Gravity, 
	University of Groningen, Nijenborgh 4, 9747 AG Groningen, The Netherlands}
\affiliation[b]{Theory Group, Department of Physics, University of Texas at Austin, Austin, TX 78712,
	USA}
\emailAdd{p.christodoulidis@rug.nl; d.roest@rug.nl; rjrosati@utexas.edu}

\abstract{
We investigate the observational signatures of many-field inflation and  present analytic expressions for the spectral index as a function of the prior. For a given prior 
we employ the central limit theorem and the horizon crossing approximation to derive universal predictions. These include a specific dependence on the prior choice for initial conditions that has not been seen in previous studies. Our main focus is on quadratic inflation, for which the initial conditions statistics decouple from those of the mass distribution, while other monomials are also briefly discussed. We verify the validity of our calculations by comparing to full numerical simulations with 100 fields using the transport method.}	

\arxivnumber{1907.08095}

\begin{document}

\maketitle

\section{Introduction}	

Inflation is the leading paradigm for the generation of primordial perturbations that seeded structure formation. The latest \textit{Planck} results confirm the generic traits of this paradigm and   are compatible with the simplest models of single field slow-roll inflation \cite{Akrami:2018odb}. In such models, the dissipative nature of inflationary dynamics reduces the order of the scalar field equation from second to first providing a notion of initial conditions independence \cite{Salopek:1990jq,Liddle:1994dx}.  This leads to specific predictions for cosmological observables, e.g.~the scalar spectral index, its running and the tensor-to-scalar ratio. For the simplest model, quadratic inflation, these are given by 
\begin{equation}
n_s  = 1 - 2/N_*\,, \qquad  \alpha_s = - 2 / N_*^2 \,, \qquad r = 8/N_*\,,
\end{equation}
where $N$ is the number of e-folds and $N_*$ corresponds to the CMB horizon crossing moment. Moreover, a common prediction of   single-field slow-roll models is that non-Gaussianities will be negligible \cite{Maldacena}.
	
Though CMB data are well described by the predictions of single-field inflation, UV-complete theories naturally predict many light scalar fields to be present in the early universe. A natural question then concerns the generalization of single-field observables to multiple fields with a non-trivial potential $V(\Phi_i)$ with $i = 1, \ldots, \mathcal{N}$ and a Cartesian metric on field space. While complicated in general, analytic control can be achieved in the slow-roll slow-turn limit, where the acceleration of every field is neglected, with approximate equations of motion
\begin{equation}\label{eq:multi-SR}
	\sqrt{3 V} \dot \Phi_i + \partial_i V \approx 0\, .
\end{equation}
However, this does not lead to a unique  dynamics for the background\footnote{This can be contrasted with models in which the interplay between the potential and field-metric can lead to one-parameter ``single-field'' trajectories, see e.g.~\cite{Christodoulidis:2019mkj}.}: eq.~\eqref{eq:multi-SR} yields an $\mathcal{N}$-dimensional hypersurface in $2\mathcal{N}$-dimensional phase space and therefore there is an $(\mathcal{N}-1)$-dimensional hypersurface that represents CMB horizon crossing. This results in an intrinsic dependence of the observables on the initial conditions, even in the slow-roll limit. Every configuration on this hypersurface will give exactly 55-60 e-folds of inflation and since there is no agreed measure on the space of initial conditions one can consider all of them to be equally probable. However, this is not the only possibility: other configurations that provide a larger number of e-folds will in general favour certain points of the CMB hypersurface more than others. The choice of the initial $(\mathcal{N}-1)$-dimensional hypersurface that gives sufficient inflation will be called the \textit{prior}.
	
Turning to perturbations, analytical expressions for an arbitrary number of fields exist in the horizon-crossing approximation\footnote{Despite its name it does not evaluate the power spectra at horizon-crossing but rather expresses observables at the end of inflation using their horizon-crossing values \cite{Frazer:2013zoa}.} \cite{Starobinsky:1986fxa,Sasaki:1995aw,Kim:2006te,Vernizzi:2006ve}. This method takes into account superhorizon evolution of curvature perturbation but ignores contributions from the fields' position at the end of inflation. Moreover, it assumes the slow-roll approximation and requires an analytic expression of $N$ in terms of the fields, and thus its applicability is more limited. On the contrary, the standard numerical approach is by means of the transport method \cite{
Dias:2015rca,Seery:2012vj,Mulryne:2010rp,Paban:2018ole,Dias:2017gva,Bjorkmo:2017nzd} which solves the perturbations' equations of motion equivalent to tree level in the in-in formalism, and requires no slow-roll or horizon crossing approximations. 

A simple and well-studied multi-field model is $\mathcal{N}$-flation, consisting of a sum of quadratic potentials \cite{Dimopoulos:2005ac, Piao:2006nm}. This model has received interest in the axion landscape community, see e.g.~\cite{Long:2016jvd,Bachlechner:2018gew,Bachlechner:2019vcb}, because it can approximate inflation towards a cosine minimum. Early investigations relied on the horizon crossing approximation which allows for simple calculation of the spectral index, tensor to scalar ratio, running and non-Gaussianity. More specifically, $r$ and the non-Gaussianities are found independent of initial conditions and the number of fields \cite{Alabidi:2005qi,Kim:2006te}, whereas the spectral index and its running inherits the dependence on initial conditions \cite{Kim:2006ys, Easther:2005zr}.

In the many-field limit ($\mathcal{N} \rightarrow \infty$), however, predictions for quadratic fields have been shown to become sharp and universal \cite{ Easther:2013rva}, while similar predictivity has been found in recent many-field numerical investigations in other contexts \cite{Price:2014ufa,Dias:2016slx,Dias:2017gva,Bjorkmo:2017nzd}\footnote{In our analysis the fields' masses are taken constant with $\mathcal{N}$, but some models (e.g. those based on random matrix theory), have taken the fields' masses to grow like $\sqrt{\mathcal{N}}$. In these cases, the $n_s$ we compute depends on ratios of masses to equal powers, so is identical in the many-field limit.}. This universality stems from the central limit theorem and in \cite{Easther:2013rva} it was shown that different priors have weak, and negligible, dependence on observables. It is the goal of this paper to investigate to what extend the results found in earlier literature are prior independent.

The paper is organized as follows: in Sec.~\ref{sec:obs_nf} we discuss observables for a sum of quadratic fields. We discuss and extend the previously used priors and derive new analytical formulas at the many-field limit, demonstrating both the universality and prior dependence of observables. In Sec.~\ref{sec:nr} we compare our analytical formulas with full numerical simulations of 100 fields using the transport method. In Sec.~\ref{sec:hm} we extend our analysis for higher monomial potentials. Finally, in Sec.~\ref{sec:concl} we offer our conclusions.

\section{Computing Observables for Many-field $\mathcal{N}$-flation} \label{sec:obs_nf}

Using the horizon-crossing approximation, the spectral index and other observable quantities are expressable in terms of the fields' values at horizon crossing.
However in order to study these observables' prior dependence, we need to express them in terms of the initial field configuration using the equations of motion.
For any sum-separable potential, the slow-roll equations of motion \eqref{eq:multi-SR} can be solved exactly. In our case \textcolor{red}{\footnote{Although $\mathcal{N}$-flation was introduced as a model with sufficiently large number of fields, here $\mathcal{N}$ is arbitrary. Moreover, we will use multi- to refer to $\mathcal{N} \approx \mathcal{O}(1)$ number of fields, while many- indicates $\mathcal{N} > \mathcal{O}(10)$.}} 
\begin{equation}
V =  \sum_i \frac12 m_i^2 \Phi_i^2 \, ,
\end{equation} 
and we obtain the solution
\begin{equation*} \label{eq:ev_nf}
\Phi_{i,*} = \Phi_{i,0} e^{-m^2_i  \tau_*} \, ,
\end{equation*}
where $ \Phi_{i,0} $ and $\Phi_{i,*}$ are the initial and horizon-crossing field displacements respectively. The time $\tau$ is defined by $\ud N = - V \ud \tau$ and ranges from $\tau=0$ to $\tau_*$ during $N=N_0$ to $N_*$.  Moreover, since the potential is sum-separable $V = \sum W_i(\phi_i)$ using the slow roll equations of motion \eqref{eq:ev_nf} the number of e-folds can be calculated from \cite{Starobinsky:1986fxa}
\begin{equation}
	N = \sum_i \int \ud \phi_i {W_i \over \partial_i W_{i}} \, .
\end{equation}
Thus, field values happen to satisfy a so-called hypersphere constraint at all times during the evolution 
\begin{equation} \label{eq:hsc}
\sum_i \Phi_{i}^2 = 4 N \, ,
\end{equation}
where the field values are evaluated at $N$ e-folds before the end of inflation.

The previous sums can be rewritten in a convenient way by introducing the \textit{sample average}. For an arbitrary random variable $A$ 
we define the sample average as 
\begin{equation}
 \langle A \rangle_s \equiv {\sum A_i \over \mathcal{N}} \, ,
\end{equation}
which is also a random variable. In the limit $\mathcal{N} \rightarrow \infty$, if the conditions of the central limit theorem are satisfied, then the sample average converges to the \textit{expectation value} $ \boldsymbol{\langle} A \boldsymbol{\rangle}$. Moreover, we will need joint expectation values of $\Phi_0$ and $m^2$ to arbitrary powers $k$ and $l$ respectively
 \begin{equation}
  \langle \Phi_0^k (m^2)^l \rangle = \int \ud \Phi_0 \ud (m^2) \,  \Phi_0^k (m^2)^l  P(\Phi_0,m^2) \,,
  \label{eq:jpdf} 
 \end{equation} 
 where $P(\Phi_0,m^2)$ is the joint distribution of fields and masses. When  the two random variables are independent, the probability distribution becomes product-separable and the average value splits into product of averages $\langle \Phi_0^k (m^2)^l \rangle = \langle \Phi_0^k \rangle \langle (m^2)^l \rangle $.  

According to the $\delta N$ formalism, if the slow-roll slow-turn relations hold then the power spectrum at the end of inflation is given by \cite{Sasaki:1995aw}
\begin{equation}
	P_R = {H^2 \over (2\pi)^2 } \partial_i N \partial^i N \Big|_{N=N_*} \, ,
\end{equation}
and for monomial potentials in general the spectral index is given by
\begin{equation}
	n_s =1 - 2\epsilon_* -{1 \over N_*} \, .
\end{equation}
It will prove convenient to rescale the initial values of the fields $\Phi_{i,0} =2 \sqrt{N_0 / \mathcal{N}} \phi_i $ to obtain the normalisation $\langle \phi^2 \rangle_s=1$. The spectral index at horizon crossing in this notation is given by 
\begin{equation} \label{ns_multi}
n_s =  1 - \frac{1}{N_*} \left(1 + \frac{\langle e^{-2m^2 \tau_*} \phi^2 \rangle_s \langle m^4 e^{-2 m^2  \tau_*} \phi^{2} \rangle_s}{ \langle m^2 e^{-2 m^2 \tau_*} \phi^2 \rangle_s^2} \right) \,,
\end{equation}
where  the sample averages are calculated for a specific realization of the fields and masses, and we used 
\begin{equation}
N_0 \langle e^{-2m^2 \tau_*} \phi^2 \rangle_s = N_* \, .
\end{equation}
For a choice of masses (drawn from a given mass distribution) the numerical value of the previous equation depends on the fields' realization, leading to an intrinsic initial conditions dependence\footnote{It can be shown that $n_s$ is strictly bounded  $n_s \in [n_{s,min}, 1-2/N_*]$, with the lower bound corresponding to a configuration where only the heaviest and lightest fields are non-zero and contribute with equal energies.}. In the following we will discuss two physically well-motivated priors that have been considered in the literature and examine how time evolution affects predictions \footnote{Hartle-Hawking based priors (e.g.~\cite{Hertog:2013mra,Hartle:2016tpo}) are also a physically well-motivated choice, and lead to motion in purely the smallest-mass direction and observables identical to single-field inflation.}. 

\textit{$N_0$-prior} \cite{Frazer:2013zoa}: Every vector $\vec{\phi}$ corresponds to a point of the $\mathcal{N}$-dimensional hypersphere with radius $1$. The method to obtain a random point includes sampling from the multi-variate distribution $N_{\rm GS}(\mathbb{0},\mathbb{1})$ and then dividing by the norm of the vector. Assuming that the $\phi_i$'s are uncorrelated with the masses and employing similar techniques as \cite{Price:2014ufa,Guo:2017vxp}, one can show that at the limit of infinite number of fields Eq.~\eqref{ns_multi} is normally distributed with mean 
\begin{equation}
n_s \xrightarrow[\mathcal{N}\to\infty]{} 1 - \frac{1}{N_*} \left(1 + \frac{\langle e^{-2 m^2  \tau_*} \rangle \langle m^4 e^{-2 m^2  \tau_*} \rangle}{ \langle m^2 e^{-2 m^2\tau_*}\rangle^2 } \right) \, ,
\end{equation}
and standard deviation that scales as $1/\sqrt{\mathcal{N}}$; thus, we find sharp many-field predictions. The expectation values are integrals over the mass distribution that should be evaluated at $\tau_*$, the latter given as a solution of the integral equation 
\begin{equation} \label{eq:taudef}
\langle e^{-2m^2 \tau_*} \rangle \equiv \int \ud m^2 P(m^2)e^{-2m^2 \tau_*} = {N_* \over N_0} \, .
\end{equation}
Thus, this choice of prior results in a spectral index that is given by a specific time-dependent combination of the moments of the mass distribution: at $\tau = 0$, i.e.~starting at a random point on $N_*$, this is given by the variance of the distribution, while starting at a random point at a higher $N_0$ there will also be higher-order moments that contribute.

\textit{$E_0$-prior} \cite{Easther:2013bga}: Instead, one can start with a fixed initial energy $E_0$, and assume the energy per field to be uncorrelated with the mass distribution. Defining initial energies as $2 E_i = m^2_i \phi_i^2$ its sample average is $\langle E \rangle =  E_0/(4N_0)$. The central limit theorem implies that, at large $\mathcal{N}$, \eqref{ns_multi} becomes
\begin{equation} 
 n_s \xrightarrow[\mathcal{N}\to\infty]{} 1 - \frac{1}{N_*} \left(1 + \frac{\langle m^{-2} e^{-2 m^2  \tau_*}\rangle \langle m^2 e^{-2 m^2  \tau_*}\rangle }{ \langle e^{-2 m^2 \tau_*}\rangle^2 } \right) \, .
\end{equation}
The initial energy and the number of e-folds are related by $E_0={2 N_0 / \langle m^{-2} \rangle}$ while $\tau_*$ is calculated by Eq.~\eqref{eq:taudef}. Hence, for this initial conditions prior, the spectral index is given by a different time-dependent combination\footnote{For quadratic inflation and these two priors (or any prior for the quantity $m^p \phi$), the distribution $P(\phi)$ is not necessary for the calculation of $\langle n_s \rangle$ since the field- or energy-dependent terms decouple from masses.} of the moments of $P(m^2)$. 

 The asymptotic behaviour (for either prior) can be inferred as follows: at sufficiently large time $\tau_* \gg 1$, the lightest field $\phi_1$ (provided its mass is nonzero) will dominante the numerator and denominator in Eq.~\eqref{ns_multi} and the ratio asymptotes to $\langle e^{-2m^2 \tau_*} \rangle_s /(\phi_1^2 e^{-2m_1^2\tau_*})$. Using Eq.~\eqref{eq:taudef} this term is equal to 1 and so $n_s\rightarrow 1-2/N_*$, the single field result. On the contrary, if the mass distribution is gapless (i.e.~the lightest field is massless) the previous ratio becomes undefined and the asymptotic value only depends on the behavior of the mass spectrum around the massless point. Precisely, for a mass spectrum with lowest order term $P(m^2) \propto m^{2 \alpha} + \mathcal{O}(m^{2(1+\alpha)})$, applying \eqref{eq:jpdf} and taking the asymptotic limit gives
\begin{equation}
n_s \xrightarrow[N_0\rightarrow\infty]{} 1 - \frac{1}{N_*}\left(2+\frac{1}{\gamma+\alpha}\right) \, ,
\end{equation}
where $\gamma=1 $ for the $N_0$-prior,  $\gamma=0$ for the $E_0$-prior and $\gamma+\alpha>0$.  

Finally, the above formalism allows for a straightforward calculation of the running, by differentiating
Eq.~\eqref{ns_multi} w.r.t. $N_*$. This results again in specific combinations of the time-dependent averages $q_k \equiv  \langle e^{-2m^2 \tau_*} m^k \phi^2 \rangle$:
 \begin{equation}
  \alpha_s = {1 \over  N_*^2q_2^3 } (q_2^3+ 2 q_0 q_2 q_4^2  - q_0^2 q_6 ) \,,
   \end{equation}
at lowest order in slow-roll.

\section{Numerical results}  \label{sec:nr}
\begin{figure}[H]
\centering
\includegraphics[width=0.7\textwidth]{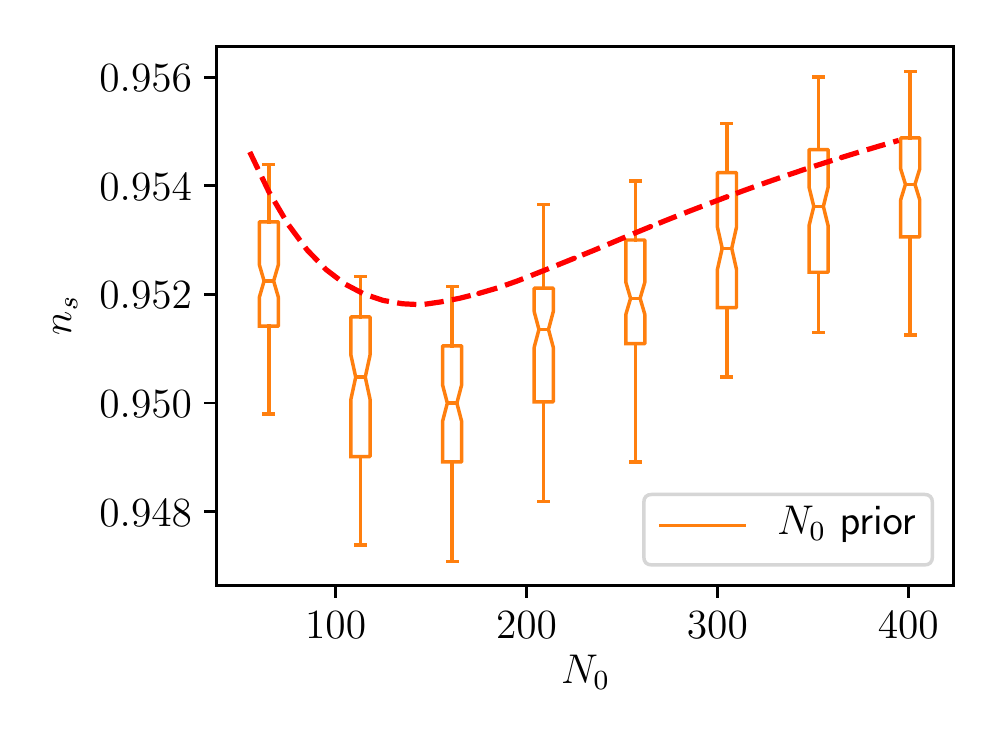}
\caption{\it{Transport method simulations of many-field $\mathcal{N}$-flation with 100 fields and initial conditions set to a fixed radius (i.e.~a fixed number of e-folds) and masses drawn from the Marcenko-Pastur distribution with $\beta=1/2$. We the compare the transport method (orange) and our analytic result (dashed, red) for initial conditions drawn uniformly over the hypersphere. At each radius, the numerical data are binned into a box and whiskers marking the 50\% and 95\% confidence intervals respectively. Agreement is at the per mil level.}}
\label{fig:numericsGapped}
\end{figure}

\begin{figure}[H]
\centering
\includegraphics[width=0.7\textwidth]{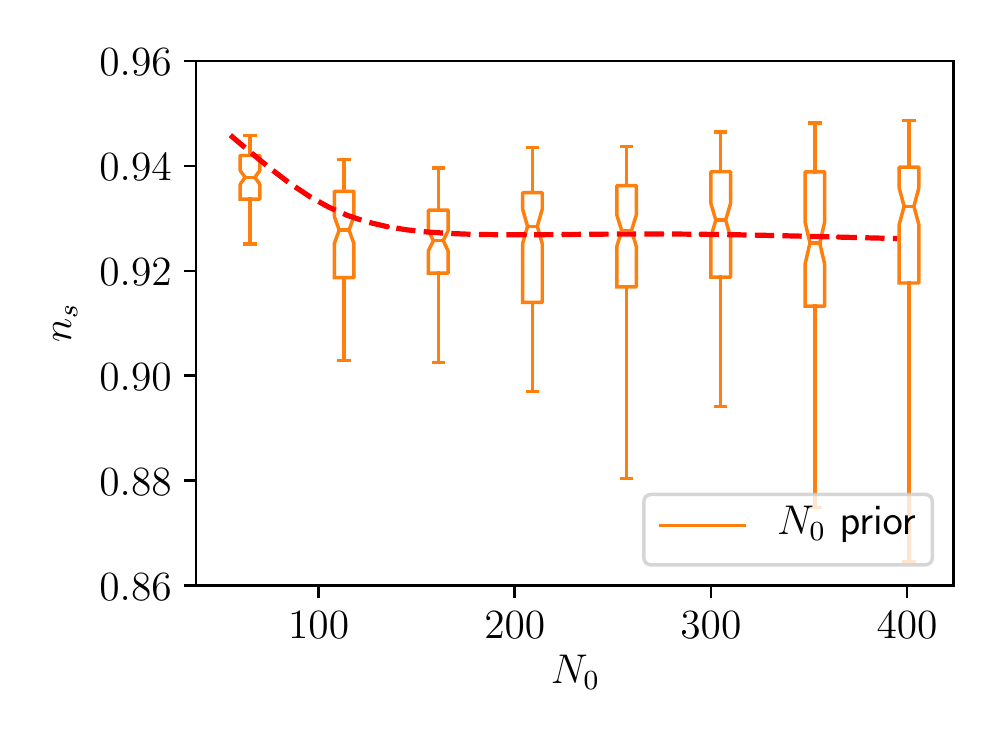}
\caption{\it{Transport method simulations of many-field $\mathcal{N}$-flation with 100 fields and initial conditions set to a fixed radius (i.e.~a fixed number of e-folds) and masses drawn from the Marcenko-Pastur distribution with $\beta=1$ (a gapless spectrum). We compare the transport method (orange) and our analytic result (dashed, red) for initial conditions drawn uniformly over the hypersphere. At each radius, the data are binned as in Fig.~\ref{fig:numericsGapped}. Agreement is at the per mil level.}}
\label{fig:numericsGapless}
\end{figure}

\begin{figure}[H]
\centering
\includegraphics[width=0.7\textwidth]{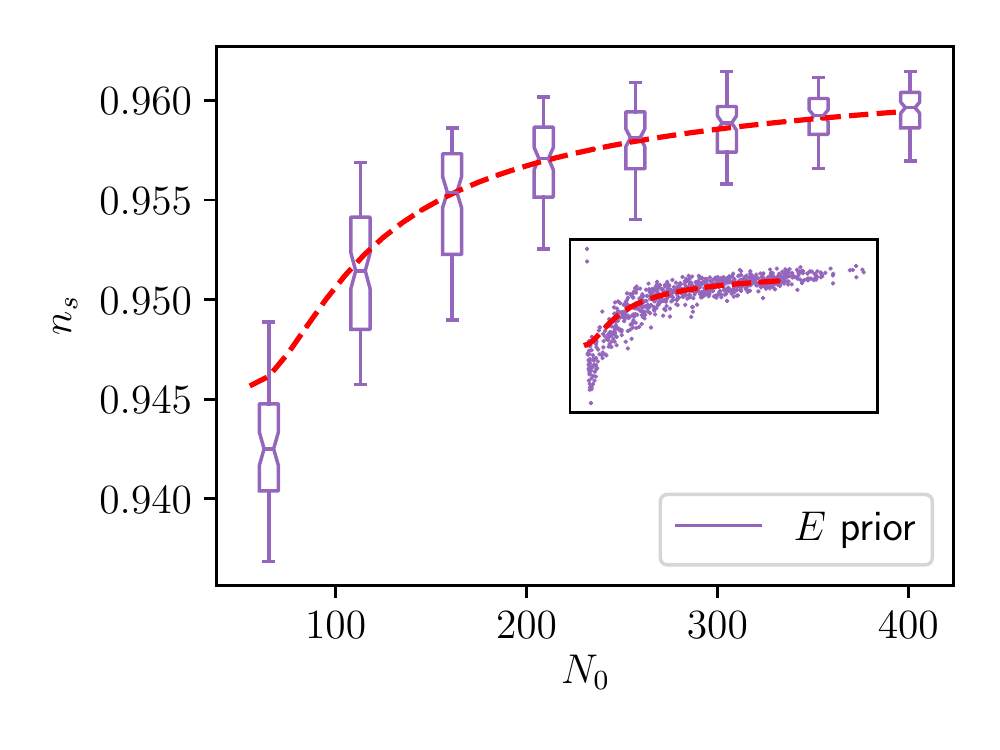}
\caption{\it{Transport method simulations of many-field $\mathcal{N}$-flation with 100 fields and initial conditions set to a fixed energy (i.e. constrained to a hyper-ellipse) and masses drawn from the Marcenko-Pastur distribution with $\beta=1/2$. The data are binned by e-folds, using the number of e-folds the realization would have if its energy were equipartioned. The number of e-folds varies by as much as $\pm 30\%$ from equipartitioned energies. At each energy, the data are binned into a box similarly to Fig~\ref{fig:numericsGapped}, or alternatively the unbinned data are displayed in the inset. The dashed red line marks our corresponding many-field analytic result.
Agreement is at least at the per mil level.}}
\label{fig:numericsEquipGapped}
\end{figure}

The above approximations and trends are confirmed by full numerical simulations. We use the \texttt{Inflation.jl} transport code\footnote{The code is written in Python and Julia and will be  publicly available soon.}, which is capable of solving the perturbations' equations of motion for $\mathcal{N} \sim 100$ using the transport method.
We take the mass distribution to be Marcenko-Pastur \cite{Easther:2005zr}
\begin{equation}
p(\lambda)=\frac{1}{2\pi\lambda \beta \sigma^2 }\sqrt{(b-\lambda)(\lambda-a)}
\end{equation} 
where $a=\sigma^2 (1-\sqrt\beta)^2$ and $b=\sigma^2 (1+\sqrt\beta)^2$. The overall normalization $\sigma^2$ drops out of the spectral index, and $\beta$ sets the width of the distribution.
With these mass distributions and a horizon crossing surface at $N_*=55$, we have taken $200$ samples per prior and plotted the spectral index in Figs.~\ref{fig:numericsGapped}- \ref{fig:numericsEquipGapped} and  \ref{fig:numericsQuartic}.

There are a number of striking results that follow from our general analysis. First of all, for a given initial hypersphere of the $N_0$-prior, the probability distribution has a clear peak in the many-field limit. Secondly, this peak value depends on the radius of the initial hypersphere. The peaked distribution was already found in \cite{Easther:2013rva} but not the dependence on the initial hypersurface. Instead, we see a clear trend: starting at $N_0=55$ the spectral index is set by the variance of the mass distribution, the peak value first goes down and reaches a minimum around $N_0 \sim 160$ for the specific mass distribution
\footnote{A minimum in $n_s$ as a function of $N_0$ will be present when $n_s(N_*)$  is lower than its asymptotic value and $d n_s / d \tau |_{\tau=\tau_*} < 0$. For a mass distribution with $\langle m^2 \rangle=1$ and an asymptotic relation $n_s \rightarrow 1-c/N_*$, these conditions imply a minimum will occur whenever $\langle m^4 \rangle > c-1$ and $\langle m^6 \rangle < 2c (c-1)$.  For Marcenko-Pastur, the expectation values for the quartic and sextic moments are given by $1+\beta$ and $1 + 3\beta + \beta^2$, respectively. The gapped case that we consider has $\beta=1/2$ and asymptotes to $c=2$ and therefore satisfies both conditions.} of Fig.~\ref{fig:numericsGapped}. Starting at yet larger radii, the heavier fields have more time to decay and this will eventually result in a single-field prediction in the large $N_0$ limit.

Next, we examine the gapless Marcenko-Pastur distribution with $\beta = 1$ in Fig.~\ref{fig:numericsGapless} for the $N_0$-prior. Instead of converging to the single-field result, we instead find $n_s \rightarrow 1-4/N_\star \sim 0.927$. This asymptotic value only depends on the behavior of the mass spectrum around the massless point $p(m^2)\sim m^{-1}$, in accordance to the discussion in the previous section.

Finally, we examine the $E_0$-prior, which corresponds to selecting random field values that have a fixed energy, forming a hyper-ellipse. When starting at CMB horizon crossing, this will lead to the spectral index determined by the moments $\langle m^{-2}\rangle$ and $\langle m^{2}\rangle$. For higher energies the resulting spectral index at CMB horizon crossing can be easily calculated from the time-dependent moments, and is illustrated in Fig.~\ref{fig:numericsEquipGapped}. We provide a similar numerical analysis that confirms the trend in this evolution.

\section{Higher monomials } \label{sec:hm}
For higher monomials of degree $n$ with 
\begin{equation}
V = \sum_{i} \frac1n \lambda_i \Phi_i^n \, ,
\end{equation} 
the number of e-folds as a function of the fields is given by $\sum_{i} \Phi_{i,0}^2 = 2 n N_0$. Specializing to $n>2$, the slow-roll equations \eqref{eq:multi-SR} imply
\begin{equation}
\Phi_{i,*} = \left(\Phi_{i,0}^{2-n} + (n-2)\lambda_i \tau_* \right)^{{1 \over 2-n}} \, ,
\end{equation}
where $\tau$ has mass dimension $2-n$. Using e.g. the $N_0$-prior introducing normalised fields $\Phi_{i,0} =2 n \sqrt{{N_0 / \mathcal{N}}} \phi_i $ and the rescaled time $\xi = \tau \left(2 n N_0/ \mathcal{N} \right)^{n/2-1}$, the moment of CMB horizon crossing is given by
\begin{equation}
\langle \left(\phi_i^{2-n} + (n-2)\lambda \xi_*  \right)^{{2 \over 2-n}} \rangle_s = {N_* \over N_0} \, .
\end{equation}
The horizon crossing formula is similarly 
\begin{equation} \label{ns_hmon}
n_s  =  1 - \frac{1}{N_*} - \frac{n}{2 N_0} {\langle \lambda^2\left(\phi_i^{2-n} + (n-2)\lambda \xi_*  \right)^{{2n-2 \over 2-n}} \rangle_s \over \langle \lambda \left(\phi_i^{2-n} + (n-2)\lambda \xi_*  \right)^{{n \over 2-n}} \rangle_s^2  }   \,.
\end{equation}
For $\mathcal{N}\rightarrow \infty$ sample averages can be calculated from expectation values if the distribution of $\phi_i$ is known.

Because of the hypersphere constraint, the distribution $P(\Phi_0)$ is not a Gaussian  for finite $\mathcal{N}$ (for instance, the distribution for one field is a sum of two delta's located at $\pm 1$). However, when $\mathcal{N}\rightarrow \infty$ the distribution $P(\Phi_0)$, which can be seen as the distribution of components of a random vector on the $\mathcal{N}$-sphere, asymptotes to a Gaussian distribution with zero mean and standard deviation $1/\sqrt{\mathcal{N}}$. For the rescaled fields $\phi$ their statistical moments\footnote{These can be calculated independently as integrals over the $\mathcal{N}$-sphere $\langle \psi_i^n \rangle = \int \ud \Omega \psi_i^n/ \int \ud \Omega$, expressing fields in spherical coordinates.} for $\mathcal{N} \rightarrow \infty$ can be reproduced by the expectation values $\langle \phi^n \rangle$ using $P(\phi)=e^{-\phi^2/2}/\sqrt{2\pi}$. With the distribution of the fields known, expectation values correspond to double integrals over fields and `masses' $\lambda$.

\begin{figure}[t]
\centering
	\includegraphics[width=0.7\textwidth]{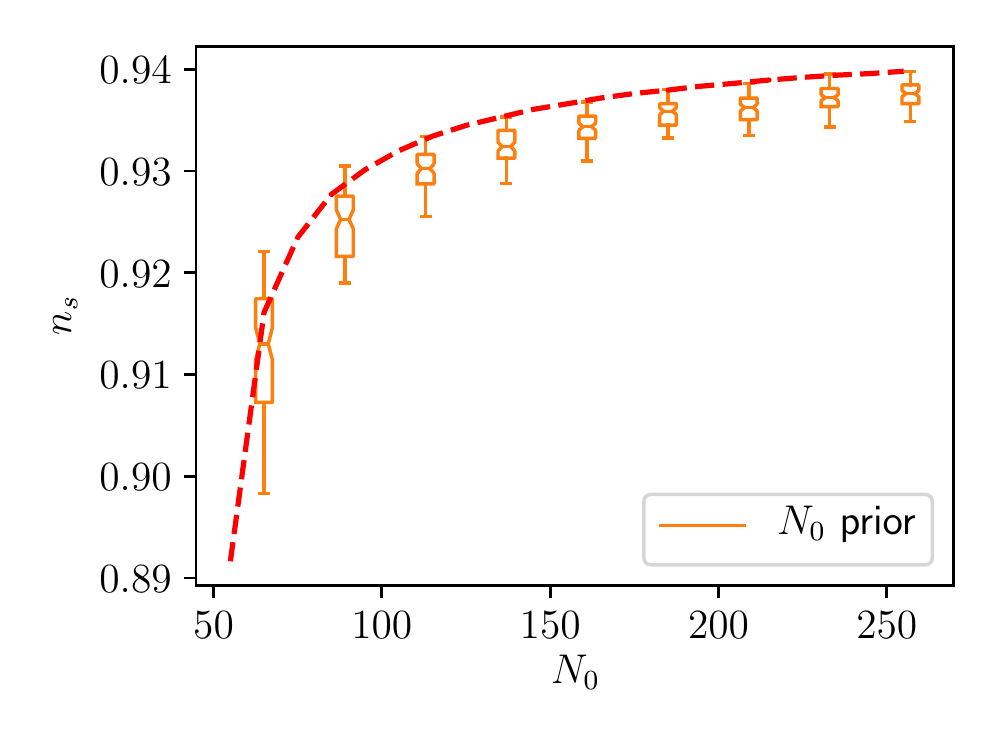}
	\caption{\it{Transport method simulations of many-field quartic monomial inflation with 100 fields and initial conditions set to a fixed radius (i.e.~a fixed number of e-folds) and `masses' drawn from the Marcenko-Pastur distribution with $\beta=1/2$. Transport numerics are in yellow and our HCA prediction in red. Agreement is at the per mil level.}}
	\label{fig:numericsQuartic}
\end{figure}

In Fig.~\ref{fig:numericsQuartic} we depict numerical simulations for quartic fields using the $N_0$-prior. In contrast to the quadratic case, there is no minimum, since the conditions for its existence are not satisfied, and instead we observe a monotonic increase towards the asymptotic value. The horizon crossing formula is again in a good agreement with the numerical results. \\

\begin{figure}[t]
\centering
\includegraphics[width=0.7\textwidth]{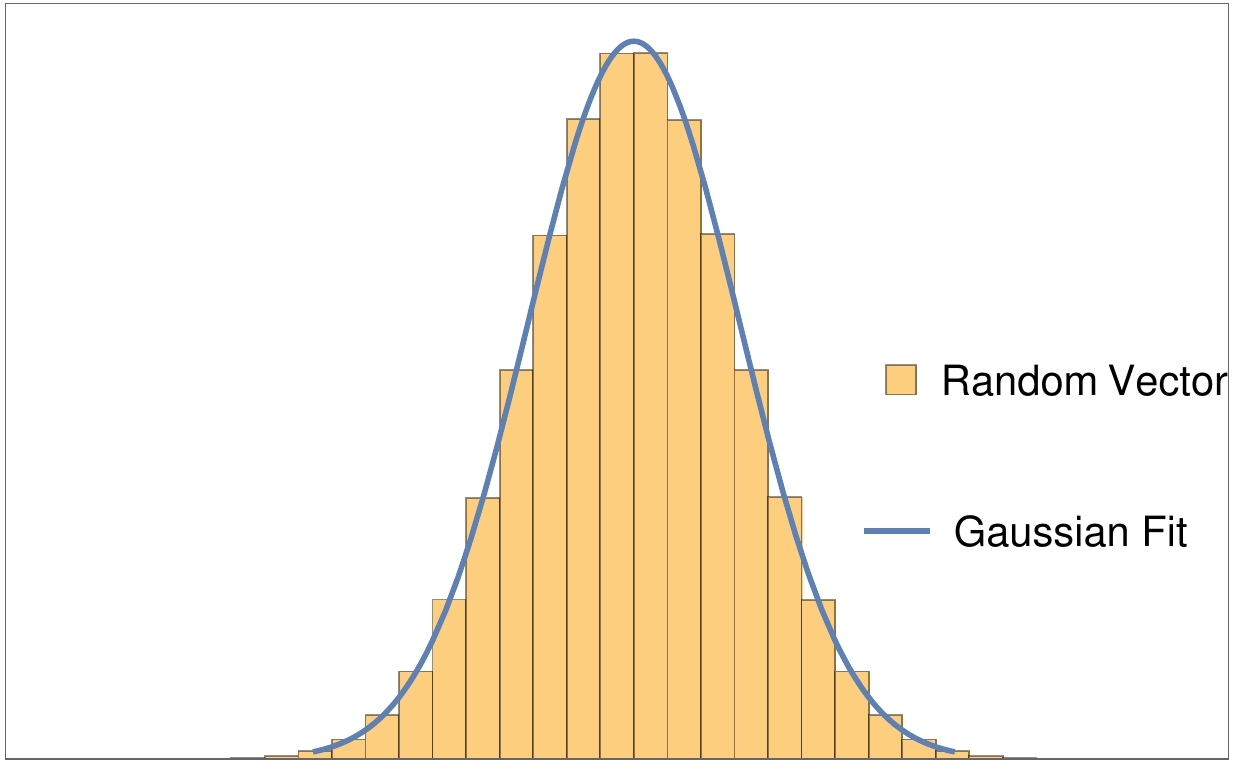}
\caption{\it{Probability density of the components of a random vector for $\mathcal{N}=10^7$ and the corresponding analytically estimated Gaussian distribution. Agreement is perfect. }}
\label{fig:gaussian fit}
\end{figure}

\section{Discussion} \label{sec:concl}
In this paper we have examined inflationary observables of sum-separable monomial potentials in the horizon crossing approximation and provided elegant analytical expressions. Although for a chosen prior there are in principle infinite different field and mass realizations (parameterized as random variables), the computed distribution of the spectral index in the many-field limit has a sharp peak. This universality can be attributed to the central limit theorem, since the analytical formulae include sums of random variables. 

While this theorem guarantees that sample averages will converge to expectation values $\langle A \rangle_s \rightarrow \int \ud \phi \ud \lambda P(\phi,\lambda)A $, the latter will depend on the joint probability distribution $P(\phi,\lambda)$, i.e.~the prior. 
Different  priors, as seen in Figs.~\ref{fig:numericsGapped}-\ref{fig:numericsEquipGapped}, can lead to different predictions, both in the spectral index and its running. Specifically, for a gapped mass spectrum, the predictions range from variance-dominated to the single-field limit. Instead, for gapless  mass distributions, the behaviour of the probability distribution for the lightest masses   determines its asymptotic behavior at high $N_0$, as seen in Fig.~\ref{fig:numericsGapless}.

We have compared our analytical predictions with numerical simulations. The excellent agreement between both approaches confirms the validity of the horizon crossing approximation. Moreover, it shows that $100$ fields suffices to reach the universal many-field regime.
	
In the absence of a non-trivial scalar-field geometry, our results for quadratic potentials can be seen as generic: the large-$\mathcal{N}$ limit pushes the horizon crossing point towards the minimum in field space, where more complicated models can be approximated with a quadratic potential. This suggests that the universality and prior dependence identified in this paper should apply to a range of more general models as well. It would be interesting to investigate the scope of our results in this direction, as well as the effects of scalar geometry in the many-field limit.

\begin{acknowledgments}
	It is a pleasure to thank Mihail Bazhba, Oliver Janssen, David Mulryne, S\'{e}bastien Renaux-Petel and John Ronayne for useful discussions and comments on the draft. We would also like to thank Jonathan Frazer, Mafalda Dias and John Stout for valuable discussions at the first stages of this project. PC and DR acknowledge support from the Dutch Organisation for Scientific Research (NWO). RR was supported by the U.S. National Science Foundation under Grant PHY-1620610.
\end{acknowledgments}
		

\end{document}